# Preparation of carbon nanotubes from graphite powder at room temperature

D. W. Lee[1,2*] and J. W. Seo[1,2]

1. Department of Physics, Cavendish laboratory, J J Thompson Avenue, Cambridge, CB3 0HE (UK)
2. 21 Nanyang Link, Physics and Applied Physics, School of Physical & Mathematical Sciences, Nanyang Technological University, 637371 (Singapore)

*Corresponding author. Fax:+65-6795-7981. E-mail address: dongwookleedl324@gmail.com (D. W. Lee)

**Abstract:**

We develop a new chemical route to prepare carbon nanotubes at room temperature. Graphite powder is immersed in a mixed solution of nitric and sulfuric acid with potassium chlorate. After heating the solution up to 70°C and leaving them in the air for 3 days, we obtained carbon nanotube bundles. This process could provide an easy and inexpensive method for the preparation of carbon nanotubes.

## 1. Introduction

After Iijima's report in 1991 [1], carbon nanotubes (CNTs), single sheets of graphene rolled into a cylinder, have been widely used and the physics of CNTs has rapidly evolved into diverse research fields: mechanics, optics, electronics and even biology. They exist in two phases: single-walled and multi-walled CNTs, with different properties. The way in which the graphene is wrapped is represented by a pair of indices (*n,m*), called the chiral vector. CNTs are classified as armchair ($n=m$), zigzag ($m=0$), or chiral (all others). Single-walled CNTs exhibit all three structures (armchair, zigzag, or chiral) while multi-walled CNTs exhibit only the first two (armchair or zigzag). CNTs have been prepared in various ways such as arc discharge [2-7], laser ablation [8-9], and chemical vapor deposition (CVD) [10-18]. CVD has proved to be the most suitable synthesis process for the production of CNTs with controlled characteristics, such as diameter, length and number of walls. However, synthesizing CNTs remains costly and difficult due to the high temperatures (around 500°C) and pressures required. Here we report on a chemical synthesis process that we have developed, which allows us to prepare CNTs easily and inexpensively at low temperatures (below 70°C) and without applying pressure.

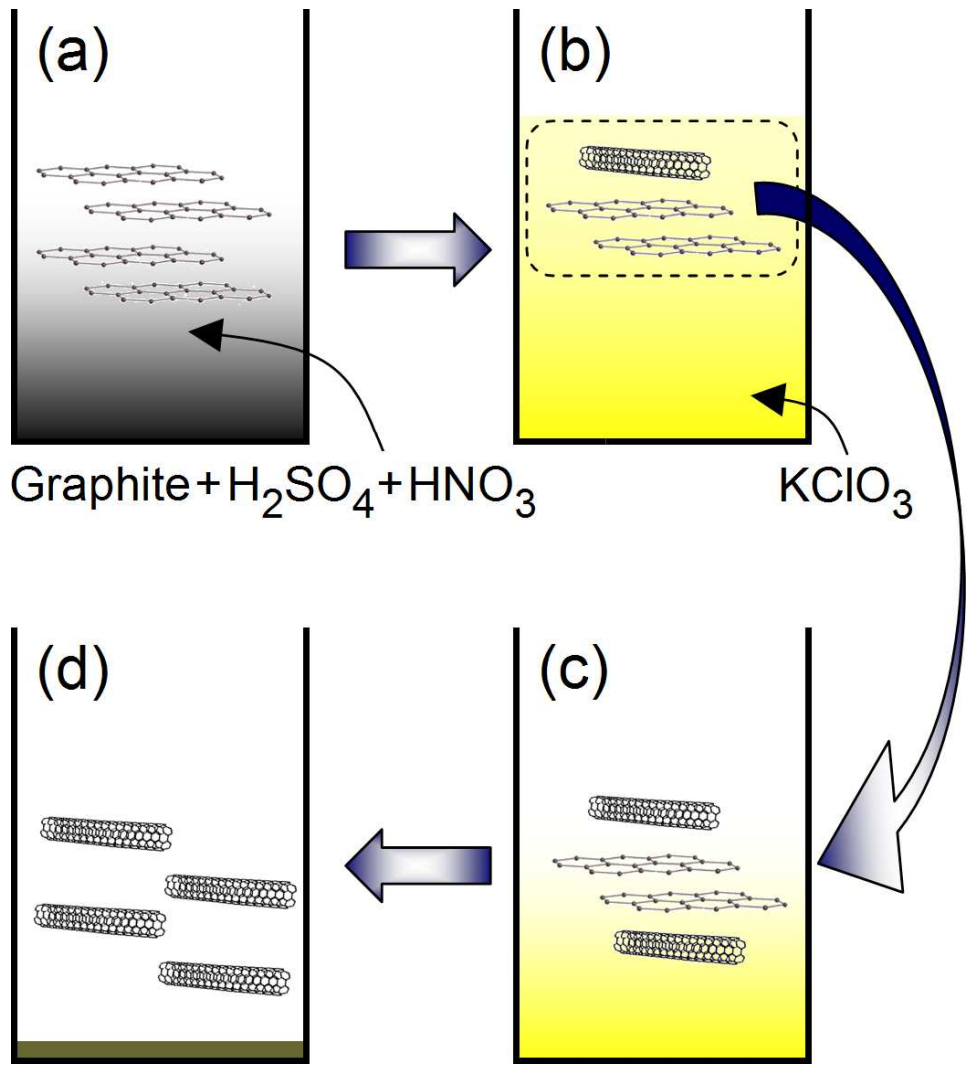

**Figure 1.** (a) Mixture of graphite, sulfuric acid ($H_2SO_4$, and nitric acid ($HNO_3$). (b) Potassium chlorate ($KClO_3$) is added to the solution. (c) Floating carbons produced from the process (b) are transferred into DI water. (d) The sample is dried after filtration. Steps (b) and (c) are repeated 4 times.

## 2. Experimental

2.1 *Sample preparation*

Our process for CNT growth is depicted in Figure 1. Our starting materials were graphite, potassium chlorate ($KClO_3$), nitric acid ($HNO_3$) and sulfuric acid ($H_2SO_4$). First, 5.0g of graphite (99.995+% purity, 45μm, Aldrich) was slowly added to a mixture of fuming nitric acid (25mℓ) and sulfuric acid (50mℓ) for 30 minutes (Figure 1(a)). After cooling the mixture down to 5°C in an ice bath, 25.0g of potassium chlorate was slowly added to the solution while stirring for 30 minutes (Figure 1 (b)). Since a lot of heat was produced while adding potassium chlorate into the mixture, we took special care during this step. The solution was heated up to 70°C for 24 hours and was then placed in the air for 3 days. Most graphite precipitated on the bottom but some reacted carbons were floating. The floating carbon materials were transferred into DI water (1ℓ) (Figure 1 (c)). After stirring it for 1 hour, the solution was immediately filtrated and the sample was dried (Figure 1 (d)). After that, the above steps (Figure 1(b) - Figure 1(c)) were repeated 4 times. The above-mentioned method for preparing CNT is similar to the Staudenmaier process [19] used for growing graphite oxide [20-22]. However, unlike the Staudenmaier process which filtrates all the graphite powders, we separate out (by filtering) only the floating graphite which has reacted with the potassium chlorate.

2.2 *Sample characterization*

To characterize the shape and structure of this sample, we used a FEI Philips XL30 sFEG Scanning Electron Microscope (SEM) and a FEI Philips Tecnai 20 Transmission Electron Microscope (TEM, operated at 200 keV).

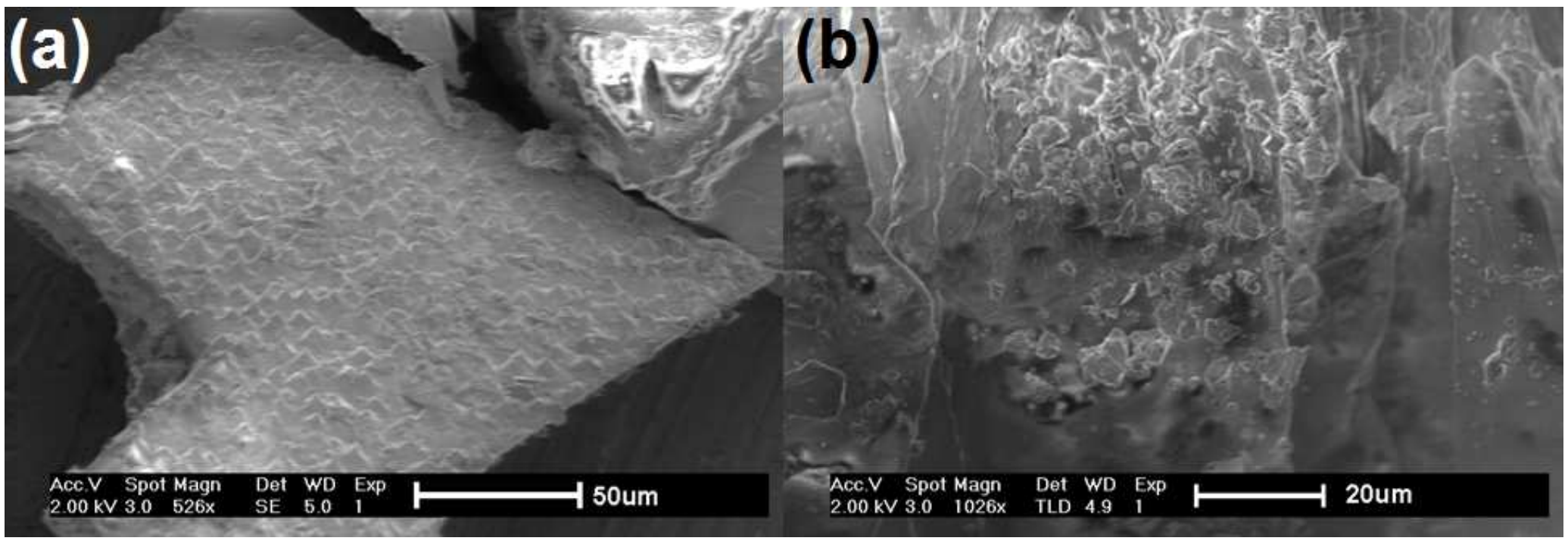

**Figure 2.** Panels (a) and (b) are SEM images of the compounds produced by the process shown in Figure 1.

**3. Results and discussion**

Figure 2(a) and (b) are SEM images of the resulting compounds. They have sharp edges and display several sheets. To investigate the morphology of the samples in more detail, TEM images were taken on a FEI Philips Tecnai 20 operated at 200 keV. The CNT bundles in the samples are clearly shown in Figure 3(a), where long and stripe-like CNTs are distinctly visible. The region indicated by number 1 in Figure 3(a) is enlarged in panel (b), revealing the whole bundles consisting of CNTs. A strained CNT pointed by an arrow in panel (b) presents its elasticity. The region denoted by the dotted circle in Figure 3(b) is magnified in Figure 3(c). The diameter of CNT is 17.01nm.

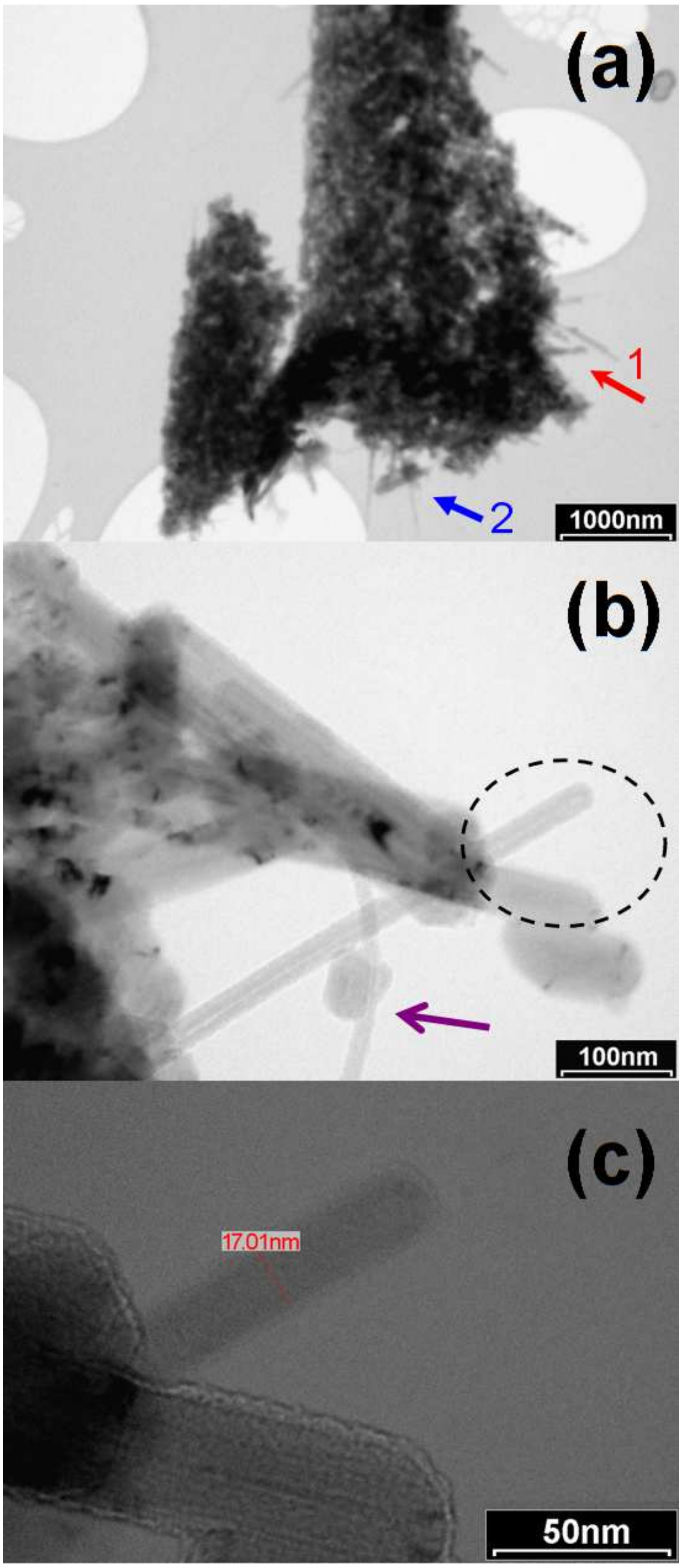

**Figure 3.** (a) TEM image of CNT bundles. (b) is an enlargement of region 1 in (a). A CNT indicated by an arrow in (b) demonstrates the CNT's flexibility. (c) enlarged region of area circled in (b), revealing a multi-walled nanotube with a diameter of 17.01nm.

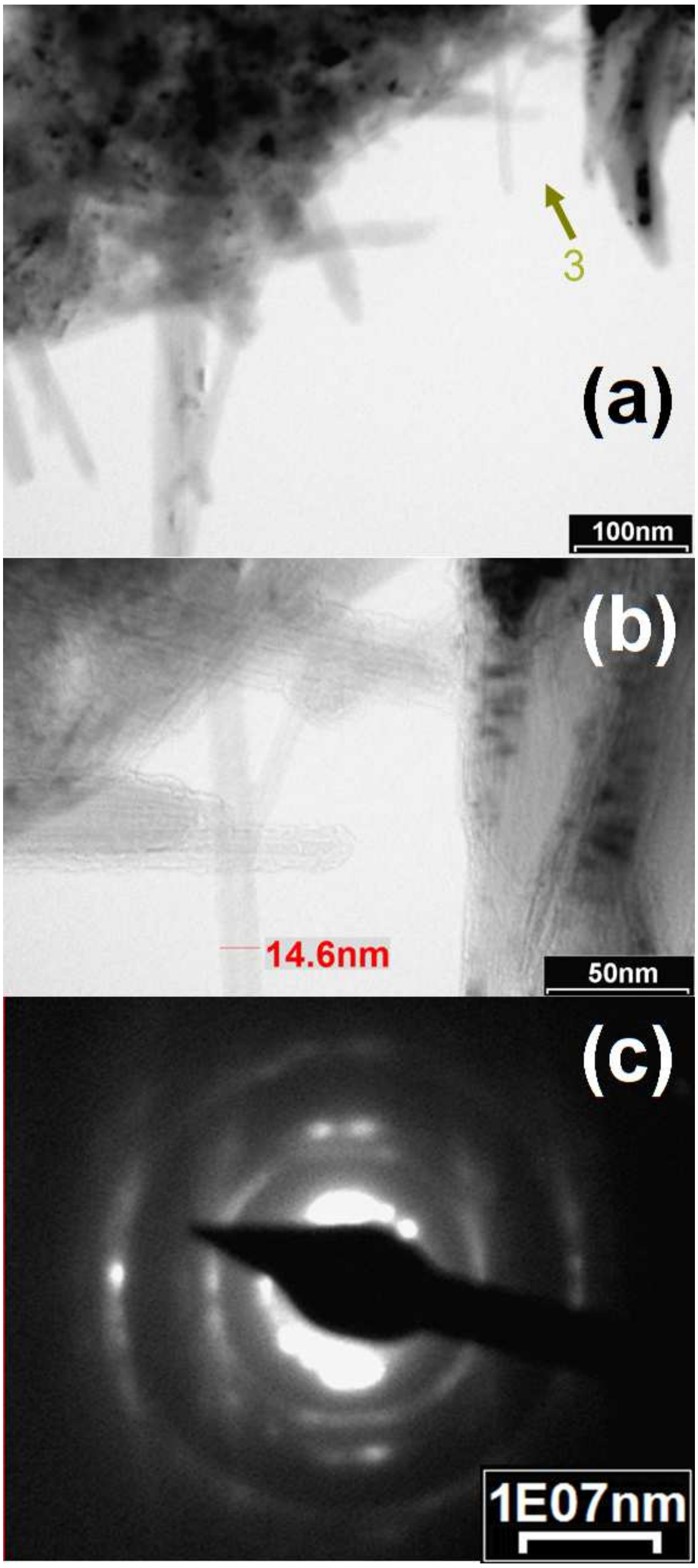

**Figure 4.** (a) presents an enlargement of region 2 in Figure 3(a). (b) is the enlarged region of (a) (arrow 3). The diameter of the multi-walled nanotube is 14.6nm. (c) Electron diffraction pattern of (b). Spots demonstrate that these CNTs contain zigzag edges and are crystallized.

Figure 4(a) presents a magnification of the region pointed out by arrow 2 in Figure 3(a). It is enlarged again in Figure 4(b), revealing that the CNTs are multi-walled. The diameter of CNT is 14.6 nm.

To determine the structure of these synthesized CNTs, electron diffraction was performed. Figure 4(c) is an electron diffraction image of the region in panel (b). The diffraction patterns look like rotation-crystal patterns of a single graphite crystal with rotation axes of (001) direction. Ring and spot patterns appear together, implying that the nanotube is comprised of both crystalline and amorphous phases. Comparison with [23] indicates that the CNTs we report here have zigzag edges.

The CNTs in the sample are randomly mixed because they were prepared by filtration after the reaction in acidic solution. Thus, they are different from CNTs which were grown on a substrate by CVD.

## 4. Conclusion

We have presented a simple chemical method for producing CNTs in liquid solution at 70 °C without any pressure treatment. The CNTs form bundles containing crystallized and multi-walled CNTs with a diameter of around 14.6nm. The electron diffraction patterns demonstrate its zigzag edge structure. We anticipate that this new synthesizing method will produce cheap CNTs and as a result allow industrial applications based on CNTs to flourish.

**Acknowledgements.** D. W. Lee is indebted to J. J. Rickard who performed the electron microscopy. J. W. Seo was supported by the Korea Research Foundation (Grant No. KRF-2005-215-C00040). We are also grateful for helpful discussion with G. R. Jelbert.

**References.**